\documentclass[final,5p,times,twocolumn]{elsarticle}
\usepackage{macros}
\usepackage{tcolorbox}
\usepackage{listings}

\usepackage{tabularx}

\begin{document}
\begin{frontmatter}
\title
{Improving Knowledge Graph Understanding with Contextual Views - Extended}
\author[wsu]{Antrea Christou}
\ead{christou.2@wright.edu}
\author[wsu]{Cogan Shimizu}
\ead{cogan.shimizu@wright.edu}
\address[wsu]{Wright State University, USA}
\begin{abstract}
    Navigating, visualizing, and discovery in graph data is frequently a difficult prospect. This is especially true for knowledge graphs (KGs), due to high number of possible labeled connections to other data. 

    However, KGs are frequently equipped with an ontology as a schema. That is, it informs how the relationships between data may be constrained. This additional information can be leveraged to improve how (knowledge) graph data can be navigated, visualized, or otherwise utilized in a discovery process.
    In this manuscript, we introduce the Interactive Knowledge (InK) Browser. This tool specifically takes advantage ontological information (i.e., \emph{knowledge}) when found in KGs. Specifically, we use modular views that provide various perspectives over the graph, including an interactive schema view, data listings based on type, neighborhood connections, and geospatial depiction (where appropriate). For this manuscript, we have evaluated the basic premise of this tool over a user group ($n=$
    With this grown user survey, we continue to evaluate how scalable tools, including flexible views, can make KG exploration easier for a range of applications.)
\end{abstract}
\end{frontmatter}
\section{Introduction}
\label{sec:intro}
Knowledge graphs (KGs) are powerful tools for organizing, linking, and integrating heterogeneous data – in both conceptualization and format \cite{kgs-hogan,kgs-hitzler} – especially when equipped with an ontology as a schema \cite{modont}. By now, they see consistent use in the public and private sectors \cite{kgs-noy}. However, their complexity often makes them challenging for users to consume \cite{thesis:rama}.

One major issue is the inherently abstract nature of the data representation. KGs consist of nodes and edges to represent entities and their relationships, which, while efficient for machines to process, can be unintuitive for users who are accustomed to more linear or hierarchical forms of data organization. Ontologies, when used as a schema for the KG, provide structure to the data, including inferrable data (e.g., through logical axioms specified in Description Logics \cite{dl-handbook}). 

Without a clear interface or visual representation, users can struggle to navigate and interpret the connections, leading to confusion and inefficiency, especially when the data and its metadata have been automatically deduced.

The specialized expertise required to understand this, as well as the dense and interconnected nature of KGs, can overwhelm users with too much information at once, making it difficult to extract (actually) relevant insights from available data. Indeed, the visualizations of graphs in general is a difficult problem, with no clear and broadly applicable solution.

On the other hand, there are significant challenges in the technical access and interaction with KGs. Often, users need specialized tools (e.g., Protégé\cite{Musen15}) or programming knowledge (e.g., RDFLib \cite{rdf-semantics}) to query and analyze the data effectively, such as understanding SPARQL  \cite{OntopSWJ} or similar query languages (e.g., cypher \cite{neo4jCypher}). For non-technical users, these requirements constitute a steep learning curve, effectively limiting the usability of KGs to experts. Even when user-friendly interfaces exist, they can risk oversimplifying the data, leading to a loss of granularity and the rich contextual insights that KGs are designed to provide. Bridging the gap between the complexity of knowledge graphs and user accessibility remains a critical hurdle to their widespread adoption. Hence, this evaluation brings us one step closer in answering the following question : \textbf{How can knowledge graphs be made easier for people with no expertise to understand and use without compromising their structural depth or semantic richness?}

In this paper, we re-introduce the Interactive Knowledge (InK) Browser, which provides a modular format for browsing a KG at multiple levels of abstraction, as well as views on the data flexible to the type of data; that is, the view will change based on context. The work herein establishes the InK Browser as both a research platform and an actual deployable tool for use in navigating and otherwise (visually) consuming data in a KG. This work expands on \cite{ink-browser} with a rigorous statistical study over a dataset of larger size and complexity than in the proof-of-concept study.

We are broadly interested in how the InK Browser can facilitate specific user tasks. It is well known that certain types of data are not easily converted to a KG format (e.g., faster data for satellite imagery) or geographic (positional) data. Instead, it is more appropriate to model the metadata and point to the data, as is, in a GIS-enabled, relational database in the case of raster imagery or in the case of positional data: a map. Adjusting to the specific type of data is an important user-driven task. Frequently, it is also the case that a user must simplify complex graphical structures into something more amenable for quick consumption. Various ontology and KG development methodologies produce graphical structures of varying complexity, which are employed in various usage scenarios. While they may accurately model a scientific domain or an explicit conceptualization of reality \cite{gruber}, it may not always be easy for a user to understand.

The Ink Browser seeks to overcome these challenges. Our study investigates two hypotheses pertaining to the use of InK Browser for a simple Question \& Answering task, which as a proxy we measure time taken to achieve a response and the accuracy of the response. Specifically, these are formulated as:

\begin{compactitem}
  \item \textbf{Hypothesis 1:} Using the InK Browser for interacting with knowledge graphs will result in faster task completion compared to traditional tools.
  \item \textbf{Hypothesis 2:} Using the InK Browser for navigating knowledge graphs will result in more accurate task completion compared to traditional tools.
\end{compactitem}

By confirming these hypotheses, the study demonstrates how the InK Browser can help users in a variety of fields by simplifying tasks, enhancing the process of decision- and simplifying access to dense KGs. The browser can be a valuable tool for knowledge graph research because of its dual focus on speed and accuracy. This helps targeted users who depend on structured data for their respective work and analyses. Quantitative measurements of accuracy and time for task completion evaluate Flexible Views' usefulness and efficacy in navigating and comprehending dense KGs. Statistical analysis on the data collected is computed, confirming the original assumptions. In addition, after task completion, we have followed up with our participants inquiring them to complete two more surveys:
\begin{compactitem}
    \item \textbf{A Posteriori Survey:} Designed to understand the participant's baseline familiarity with topics related to knowledge graphs and/or tools used along with. 
    \item \textbf{Usability Questionnaire:} Designed to assess the participant's experience using the InK Browser and how easy and efficient it is for completing tasks related to knowledge graphs.
\end{compactitem}

Concretely, our manuscript offers two primary contributions:
\begin{compactitem}
    \item A larger and more complex dataset and participant pool were used in this statistically significant evaluation of the InK Browser, which confirmed substantial advances in task accuracy and completion time.
    \item An analysis of user experience using Usability questionnaires and a posteriori surveys, providing insight into user familiarity, overall usability, and the tool's usefulness for interacting with KGs.
\end{compactitem}

The rest of this paper is organized as follows:
\begin{inparaitem}[]
    \item \textbf{Section 2} introduces the InK Browser, a user-friendly application for dynamic, schema-driven representations of complex knowledge graphs.
    \item \textbf{Section 3} frames the InK Browser within relevant work on KG usability and modular ontology design, which is supported by a framework of semantic web concepts, graph theory, and Human Computer Interaction
    \item \textbf{Section 4} covers the design, data cleaning, statistical analysis, and usability surveys of a user study that compares task performance with and without the InK Browser.
    \item \textbf{Section 5} provides statistical proof that users of the InK Browser completed tasks more quickly and accurately.
    \item \textbf{Section 6} explains how the tool enhanced performance and user experience based on survey and quantitative data.
    \item \textbf{Section 7} verifies the effectiveness of the InK Browser and outlines next steps for integrating educational videos for interactive learning.
\end{inparaitem}

\section{The Interactive Knowledge Browser}
\label{sec:ink}
By making it possible to explore RDF-based KGs in an accessible and interactive manner, the InK Browser is a web-based application that helps users who are not familiar with SPARQL overcome technical challenges.   It contributes to expanding access to semantic web technologies for a broader audience, including professionals, educators, and students.
\begin{figure*}
    \centering
    \includegraphics[width=1.0\linewidth]{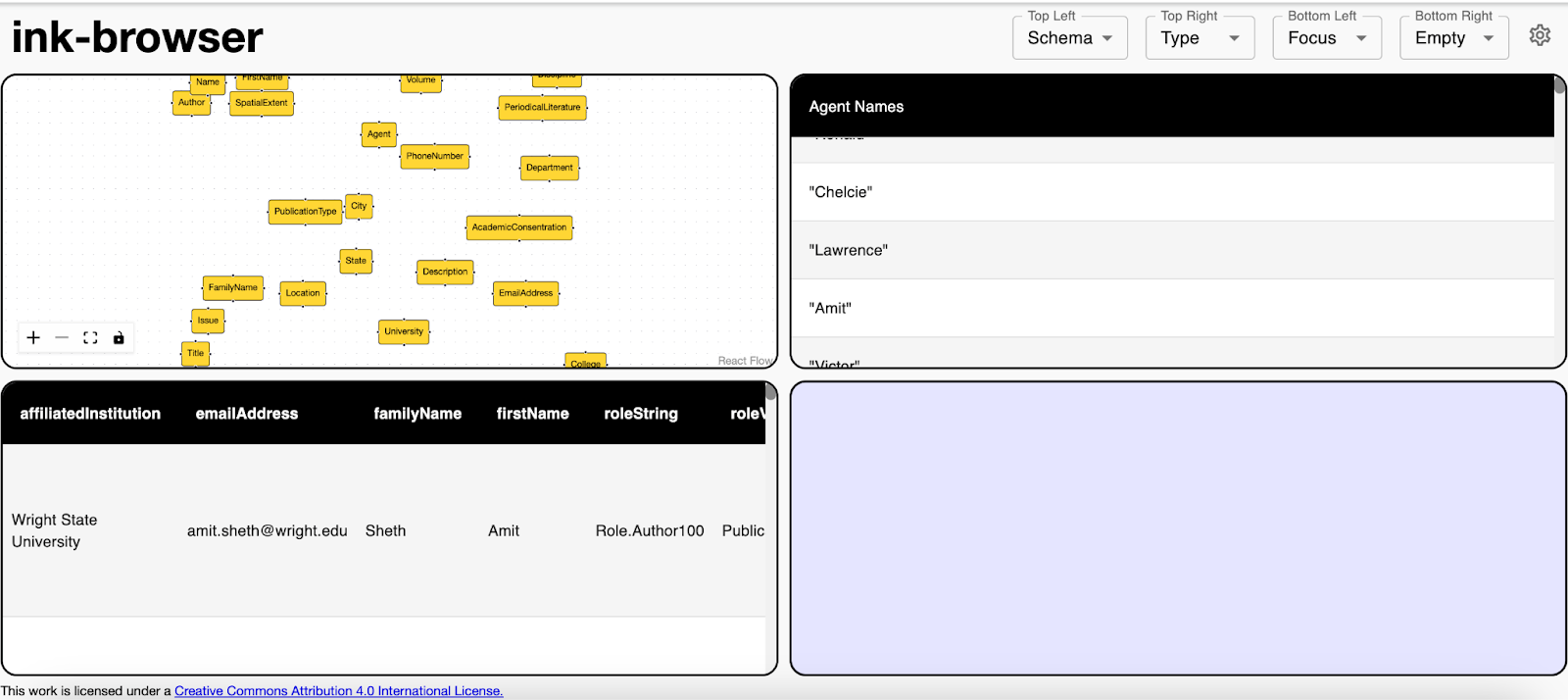}
    \caption{This figure shows the basic user interface for the InK Browser. It is separated into four, customizable quadrants. Top Left: Schema Component View. Top Right: Extracted attributes from selected component. Bottom Left: Extracted information for selected attribute in a shortcut version. }
    \label{fig:windows}
\end{figure*}
\begin{figure*}[!t]
\centering
\begin{lstlisting}
@prefix kastle-lab: <http://kastle-lab.org/>.
<http://kastle-lab.org/ontology/Agent-Role> a owl:ontology .

kastle-lab:Agent1.Chelcie a momo:Agent ;
    rdfs:label "Chelcie" ;
    kastle-lab:affiliatedWithInstitution 
    "Wright State University - Main Campus"^^xsd:string ;
    kastle-lab:hasFamilyName "Hinders"^^xsd:string ;
    kastle-lab:hasFirstName "Chelcie"^^xsd:string ;
    kastle-lab:isA "Agent"^^xsd:string ;
    kastle-lab:performsRole "Role.Author1"^^xsd:string .
\end{lstlisting}
\caption{An excerpt from the CoreScholar TTL file.}
\label{fig:ttl}
\end{figure*}

\subsection{Motivation}
\label{ssec:motivation}
The purpose of the InK Browser is to make knowledge graph (KG) exploration and analysis easier for a variety of users, including those who may lack a comprehensive understanding of ontology engineering. In contrast with standard tools, which often require an extensive knowledge of KG structures or SPARQL queries, the InK Browser offers an easy-to-use interface for dynamically exploring and retrieving relevant information.

We understand that individuals often need to be able navigate complicated datasets in an interactive and flexible manner. For rich KGs, which frequently contain complex, multi-layered facts, this is true, with traditional exploration tools like going through raw data can lead to inaccuracy and take significantly more time. Tools that enable users to visualize the structure and content of the data while decreasing their focus are helpful.   Through the utilization of Flexible Views, the InK Browser lets users interact with the dataset in a dynamic manner, customizing their exploration based on what they want in terms of analysis or simply learning regarding a specific domain.

The InK Browser makes it easier to explore in the following steps:
After loading the KG's materialized ttl files into Apache Jena Fuseki, users may run SPARQL queries on the triplestore utilizing a React-based interface without needing the data to be stored within the React application itself.

Flexible Views change dynamically as users specify constraints or concentrate their focus, displaying the data in an understandable and visually appealing way without requiring users to become proficient in querying the KG themselves.

\section{Design of the InK Browser}
\label{ssec:design}
The InK Browser requires Apache Jena Fuseki and a React application (React App) to run simultaneously. The React App, or user interface, is composed of several JavaScript (.js) files.In order to get relevant data from the uploaded materialized KG that remains constant within the triplestore, these files interact with a SPARQL endpoint in Apache Jena Fuseki. For the Flexible Views to display the modified content, the data does not have to be hosted inside the React application. Users can interact with the dataset by using the React App, which is easily accessed at localhost:3000, to run SPARQL queries to retrieve what they need from the KG while Jena Fuseki is operating at localhost:3030.
Figures 3 and 4 represent the schemas used for the materialization and dynamic query creation respectively. Both of them represent the CoreScholar data, with the rich schema showcasing the relationships in detail and the shallow one capturing the shortcut versions of the connections.

Each individual can create  personalized data representations that correspond to the underlying structure by dynamically applying complex constraint definitions using Flexible Views, a capability that is unique to the InK Browser. This study expands Flexible Views' capabilities to tackle the difficulties posed by rich KGs. The InK Browser now offers an interactive, multi-layered exploring environment through the use of dynamically created SPARQL searches and schema shortcuts  across a set of Schema, Type, and Focus windows, users may go across KGs and slowly refine their focus to extract clear, targeted information about things. This extended study helps users better understand dense datasets.

\section{Related Work}
\label{sec:rel}
To improve navigation and understanding of knowledge graphs (KGs), the work in "Improving Knowledge Graph Understanding with Contextual Views" presents flexible, schema-driven views in the InK Browser. It uses a controlled study to assess how the browser affects user comprehension. Our work builds on this by expanding the study's participant demographic and using the method on more intricate, highly connected KGs, thereby confirming the tool's efficacy in real-world circumstances \cite{christou2024}.

To increase clarity, collaboration, and scalability when building complicated ontologies, “Modular Graphical Ontology Engineering Evaluated”, emphasizes the benefits of modular techniques in ontology engineering. Users are more likely to focus on certain elements while preserving an organized overall structure when ontologies are divided into smaller, more manageable modules. In addition to improving usability, this modularity reduces the level of thinking that comes with complex and extensive ontologies. In this research, we incorporate flexible views derived from schemas into the InK Browser, leveraging this modular approach to materialized KG exploration. These views make use of modularity to offer customized, contextualized viewpoints of intricate KGs, making it easier for users to browse and interpret complicated data\cite{modular-evaluated}. 

\subsection{Theoretical framework}
\label{ssec:framework}
Using knowledge graph theory as a basis, the study investigates more complex graph structures with multiple levels, components, and connections. In order to handle the complexity of bigger, diverse datasets, the framework additionally uses contextualization concepts, which emphasize how comprehension of graph data changes with various user scenarios.   As the study looks at how different user groups with different levels of knowledge interact with challenging KGs, the importance of incorporating human-computer interaction (HCI) principles increases. The work intends to create more advanced visualization methods that assist users in navigating and interpreting the knowledge graph's increased complexity by leveraging on data visualization and semantic web theory.

CoreScholar \cite{corescholar} is the dataset selected for this study's materialization; it contains extensive information about researchers, such as their names, email addresses, and publication history. Publication subjects, authorship data, and associated metadata are also included in this.
 
The AgentRole pattern from MODL (Modular Ontology Design Library) \cite{modl} is used in the materialization of CoreScholar. This pattern is particularly well-suited for presenting entities such as agents and their roles in an organized manner. Modeling the relationships between researchers (agents) and their tasks in a variety of cases, including authorship, contact information, role providers etc, is made easier by this pattern. The dataset materialization ensures that the complex relationships between agents and their responsibilities are carefully recorded by following this design approach that assures modularity throughout the KG materialization and querying process.

\section{Methodology}
\label{sec:method}
The InK Browser's capability is expanded in this study to assess how well it navigates dense and highly connected knowledge graphs (KGs). The process involves  sophisticated flexible views that produce contextualized SPARQL queries and schema-driven summaries automatically.   These features make it easier for users to understand and navigate the KG's complex relationships. A 
Shallow Schema Diagram of the CoreScholar dataset which was used for the dynamic creations of the SPARQL queries.

\begin{figure*}[t]
    \centering
    \includegraphics[width=1.0\linewidth]{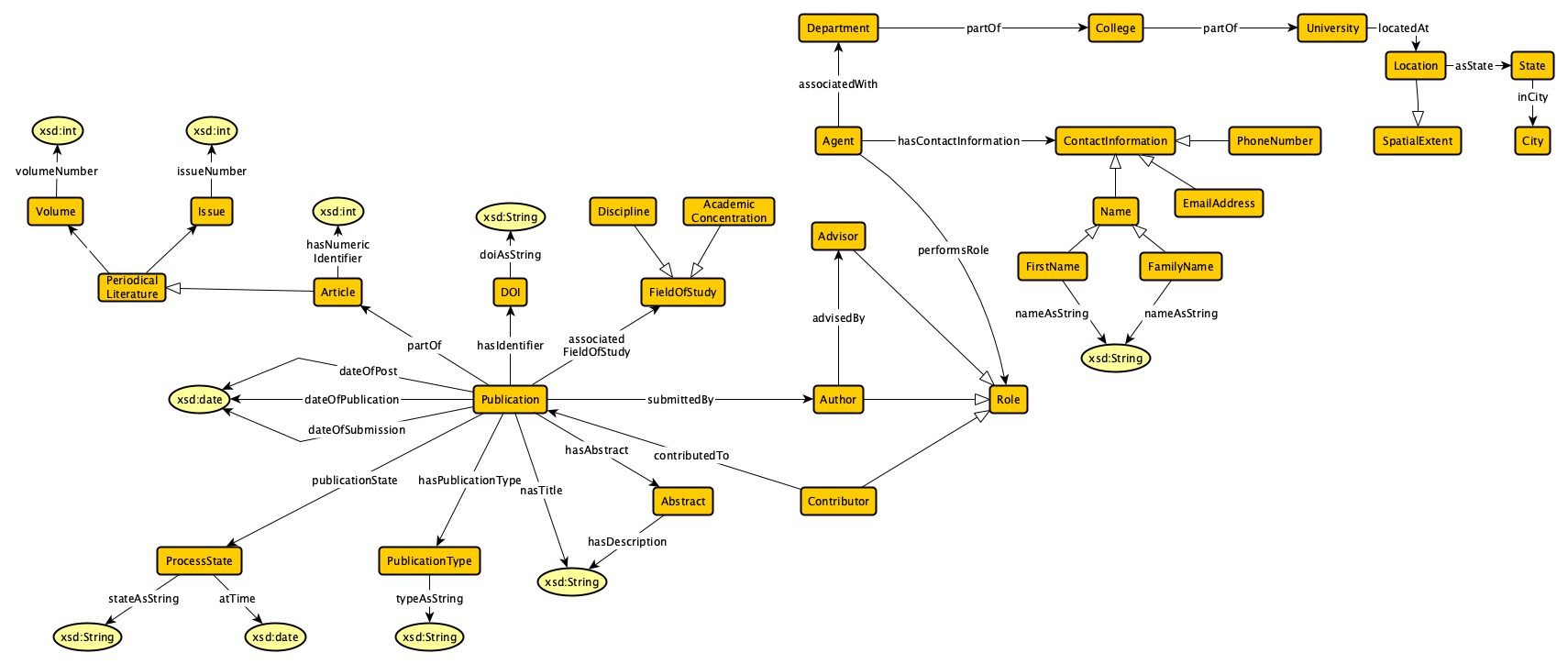}
    \caption{Rich Schema diagram of the CoreScholar dataset that the KG was materialized upon.
}
    \label{fig:core-scholar}
\end{figure*}

\begin{figure*}[t]
    \centering
    \includegraphics[width=1.0\linewidth]{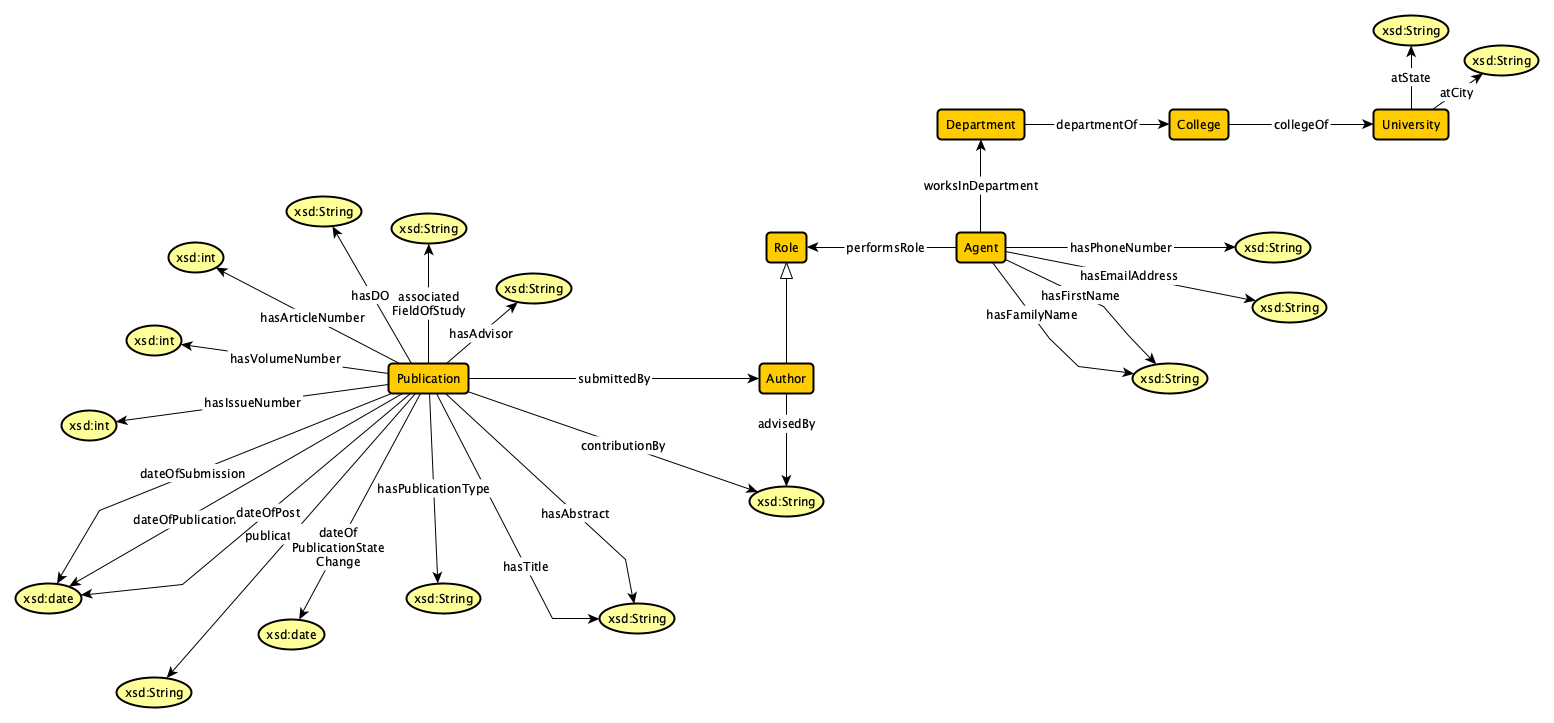}
    \caption{Shallow Schema Diagram of the CoreScholar dataset which was used for the dynamic creations of the SPARQL queries.}
    \label{fig:core-scholar-shallow}
\end{figure*}

User research was carried out with a larger participant pool and a more challenging dataset to validate the tool's effectiveness. With and without the tool, participants completed tasks that required accuracy with the time being collected as well, allowing completion of tasks metrics to be quantitatively compared. With an emphasis on improved task accuracy and completion time, statistical analysis, including t-tests and correlation analysis, was utilized to assess how tool usage affected performance.

\begin{figure}
    \begin{framed}
    \begin{enumerate}
      \item What is Amit's Family Name?
      \item Has Mauricio got any publications in the last 5 years?
      \item What is Catherine's email?
      \item What is Shelley's most recent publication field?
    \end{enumerate}
    \end{framed}
    \caption{Task example questions.}
    \label{fig:questions}
\end{figure}

\subsection{Research Design}
\label{ssec:res-design}
The first step in the experimental design was for the computer science department to recruit volunteers by email and set up a specific day and time for them to attend our lab. The first person was randomly assigned to begin Task A with the tool or without upon arrival, and this choice dictated the assignment of the following participants. This meant that if the first person used the tool, the second person had to do the same thing without it, and vice versa, with each participant taking turns. There were two sets of the assignments, each with four questions as seen from an example in picture 5. One set of questions was answered by looking through a raw TTL file, and the other set was answered while using the InkBrowser application. Participants completed each set, then used a Google survey to complete and submit their responses.   The participants' efficiency in each condition was measured by carefully tracking the amount of time needed to finish each task throughout the process. For every task, participants' accuracy was scored on a 0–4 scale, where 0 denoted no right answers, 1 denoted one right answer, and so on up to 4 right answers. 

Participants were prompted to fill out  two confidential follow-up surveys after completing the experiment.   Using a System Usability Scale (SUS) questionnaire \cite{brooke1996sus}, the initial survey evaluated the InK Browser's usability by measuring user happiness, confidence, and ease of use. Ten Likert-scale \cite{likert1932technique} survey questions about the InK Browser and raw dataset user experience were included. The second, a posteriori questionnaire collected information on participants' past knowledge and familiarity with the InK Browser and knowledge graphs using five point Likert-scale questions \cite{digitalgov_usability}.

\subsection{Evaluation Strategy}
\label{ssec:eval-strat}
Our study's objective is to assess how the InK Browser, a web-based tool that is modular and designed for knowledge graph exploration, improves user understanding and navigation of structured data. To provide a comparison within the same topic, participants completed two sets of tasks: one with and one without the InK Browser. Participants alternately started with the tool or the raw dataset for each task, which consists of answering a series of questions based on a dense KG. Time-to-completion and answer accuracy have been measured for each question. Participants used a conventional file viewer to engage directly with the raw data during the tasks, and they used a cursor for interacting with the tool. They filled out a Google Form with their responses.

Participants were invited to fill out two confidential follow-up surveys after their session. With a focus on ease of use of use, confidence, and user satisfaction, the first follow up survey aimed at evaluating the InK Browser's usability and user experience. In the second, participants' prior knowledge and familiarity with knowledge graphs were gathered. To standardize replies, both surveys employed a Likert scale method.   To ensure that no identifying information was connected to the results, all survey replies were kept private and unrelated to the initial task performance data.
\subsubsection{Data Cleaning}
\label{sssec:cleaning}
When it came to our task related responses, we started with exporting the form data into a CSV file, which included all user responses, in order to gather and organize responses from Google Form submissions. To guarantee consistency and practicality, the data was then combined into a single file and a binary column was added to differentiate the data that was collected when the tool was used and when it was not used. Then we moved on to the cleaning and the preparation stages. To do this, time-related data needed to be transformed into a seconds in order to be standardized. Then, the data was grouped by tool usage in order for us to observe those differences accuracy and task completion time has with respect to that.

For ouf follow up responses, in order to focus on participant responses, we first removed the timestamp columns that were not necessary for our analysis. The responses were later normalized by converting Likert-scale answers to numerical values, guaranteeing uniformity across the a posteriori and SUS surveys.
Specifically we assigned:
"Strongly Disagree": 1,
    "Disagree": 2,
    "Neither Agree nor Disagree": 3,
    "Agree": 4,
    "Strongly Agree": 5
The dataset was kept consistent by ignoring any responses that were unanswered since empty feedback adds no value to our insights.

\subsubsection{Statistical Analysis}

By conducting a number of tests, we compared task performance accuracy with and without the use of tools in our statistical analysis. For both groups, we first calculated summary statistics, such as the mean, variance, minimum, and maximum task accuracy levels  with and without tool usage. This gave a first understanding of the distribution of the data. The assumption that the variances of the two groups were identical was then evaluated, which is crucial for choosing the right statistical test. Independent and paired t-tests \cite{student-test} were used to evaluate the accuracy differences between the two cases.   The independent t-test, which assumed unequal variances, tested if there was a significant difference between the two groups, whereas the paired t-test evaluated the same hypothesis for matched pairs of data. We followed the same approach for the time related data.

\subsubsection{Usability Questionaire Analysis}

Ten questions, each with a 5-point Likert scale score, formed up the Usability questionnaire. These questions covered topics such the InK Browser's usability, the confidence users have in using the tool, and their opinions on its effectiveness.   A correlation matrix for the first nine questions (Q1-Q9) was produced as part of the data analysis. Any connections between various elements of the user experience, such as tool familiarity and simplicity of use, were found using the correlation matrix. To illustrate the connections between the questions, a heatmap was created, offering a thorough understanding of the relationships between the answers on different questions.
Additionally, pie charts were used to display the response distributions for every question. The data visualizations made it easier to see how people rated the various usability features of the InK Browser. The quantitative data was given further context by an investigation of open-ended responses (Q10), which made it possible to identify commonalities like the requirement for technical assistance and training.   A word frequency analysis was performed to measure the frequency of terms used in open-ended responses.

\subsubsection{A Posteriori Analysis}

In order to learn more about participants' prior knowledge and expertise with knowledge graphs, SPARQL, and the InK Browser, the A Posteriori questionnaire had five questions. The results were examined using a correlation matrix to look into connections between users' perceptions of the InK Browser's usability and their prior understanding of knowledge graphs and SPARQL. With the implementation of this matrix, patterns in the ways that participants' ratings of the InK Browser were impacted by their past knowledge and experience with knowledge graphs can be found.
The distribution of answers to each of the five questions was also shown using pie charts. This gave a brief overview of the participants' self-reported knowledge graph, SPARQL, and InK Browser expertise as well as their level of comfort using knowledge graph techniques in real-world situations.

\subsubsection{Data Preparation}
\label{sssec:data}
The process started with exporting the form data into a CSV file, which included all user responses, in order to gather and organize responses from Google Form submissions. To guarantee consistency and practicality, the data was then combined into a single file before moving on to the cleaning and the preparation stages. In order to do this, it was necessary to  transform time-related data into a common format (such as converting minutes to seconds), and make sure that every column had the proper title. After that, the data was organized into pertinent groups, like tool usage, and the amount of time spent on every task was recorded in a suitable manner.

In order to focus on participant responses, we first removed columns that weren't necessary, including timestamps, from the follow-up survey data. The responses were later normalized by converting Likert-scale answers to numerical values, guaranteeing uniformity across the a posteriori and SUS surveys. The dataset was kept consistent by replacing any missing values with zeros.

\subsubsection{Statistical Analysis}
\label{sssec:stats}
By conducting a number of tests, we compared task performance accuracy with and without the use of tools in our statistical analysis. For both groups, we first calculated summary statistics, such as the mean, variance, minimum, and maximum task accuracy levels  with and without tool usage. This gave a first understanding of the distribution of the data. The assumption that the variances of the two groups were identical was then evaluated, which is crucial for choosing the right statistical test. Independent and paired t-tests were used to evaluate the accuracy differences between the two cases.   The independent t-test, which assumed unequal variances, tested if there was a significant difference between the two groups, whereas the paired t-test evaluated the same hypothesis for matched pairs of data. We followed the same approach for the time related data.

\section{Results}
\label{sec:results}

\subsection{Statistical Analysis Results}

Tables \ref{tab:tool-usage-summary-accuracy} and \ref{tab:tool-usage-summary-time} below report basic statistics of min max mean and variance computations of the data collected when grouped by tool usage in order to get an insight of the overall behaviour of our data. Also tables \ref{tab:accuracy-test} and \ref{tab:time-test} showcase the computations of the test statistic and p-test numbers of mentioned grouped data.
The summary statistics for task accuracy in table 1 show that when the subjects used the tool, it resulted in better mean accuracy (mean = 3.6364, variation = 0.4132) than those who did not (mean = 1.7273, variance = 0.8347). A substantial difference in task accuracy between the two groups is suggested by an independent t-test (T-statistic = 7.8314, p < 0.0001), and this conclusion is supported by a paired t-test (T-statistic = 9.2176, p < 0.0001) as seen in table 3. A difference in task accuracy is also strongly supported by the Wilcoxon \cite{wilcoxon} test, which likewise produces significant results (statistic = 0.0000, p < 0.0001).
On average, individuals who used the tool finished tasks far faster than those who did not (mean = 21661.1364, variance = 2046547303.39) (mean = 461.7727, variance = 38339.27) shown in table 2. A statistically significant decrease in task completion time when the tool was used by the subjects is demonstrated by the Wilcoxon test (statistic = 51.0000, p < 0.05), paired t-test (T-statistic = 2.1473, p < 0.05), and independent t-test (T-statistic = 2.1474, p < 0.05) displayed in table 4.

\begin{table}[t]
\centering
\caption{Summary statistics for task completion time grouped by tool usage}
\scriptsize
\begin{tabularx}{\columnwidth}{c|X|X}
\textbf{Metric} & \textbf{Tool Usage Time (s)} & \textbf{No Tool Usage Time (s)} \\
\hline
Min & 205.00 & 147.00 \\
\hline
Max & 848.00 & 132720.00 \\
\hline
Mean & 461.78 & 21661.14 \\
\hline
Var & 38339.27 & 2046547303.40 \\
\end{tabularx}
\label{tab:tool-usage-summary-time}
\end{table}

\begin{table}[t]
\centering
\caption{Summary statistics for task accuracy grouped by tool usage}
\scriptsize
\begin{tabularx}{\columnwidth}{c|X|X}

\textbf{Metric} & \textbf{Task Accuracy with Tool
} & \textbf{Task Accuracy with no Tool
} \\
\hline
Min & 2.00 & 0.00 \\
\hline
Max & 4.00 & 3.00 \\
\hline
Mean & 3.63 & 1.73 \\
\hline
Var & 0.41 & 0.83 \\
\end{tabularx}
\label{tab:tool-usage-summary-accuracy}
\end{table}

\begin{table}[t]
\centering
\caption{T-Statistic and P-value computed for Accuracy.}
\scriptsize
\begin{tabularx}{\columnwidth}{c|X|X|X}

\textbf{Test for Accuracy} & \textbf{Independent} & \textbf{Paired} & \textbf{Wilcoxon}\\
\hline
T-Statistic & 7.83 & 9.22 & 0.00 \\
\hline
P-Value &2.0055e-09
 & 7.9111e-09 & 4.8259e-05 \\
\end{tabularx}
\label{tab:accuracy-test}
\end{table}

\begin{table}[t]
\centering
\caption{T-Statistic and P-value computed for Time.}
\scriptsize
\begin{tabularx}{\columnwidth}{c|X|X|X}

\textbf{Test for Time} & \textbf{Independent} & \textbf{Paired} & \textbf{Wilcoxon}\\
\hline
T-Statistic & 2.15 & 2.15& 51.00 \\
\hline
P-Value &0.04

 & 0.04 & 0.01
 \\
\end{tabularx}
\label{tab:time-test}
\end{table}

\subsection{A Posteriori Survey Results}

\begin{figure*}[t]
    \centering
    \includegraphics[width=0.7\linewidth]{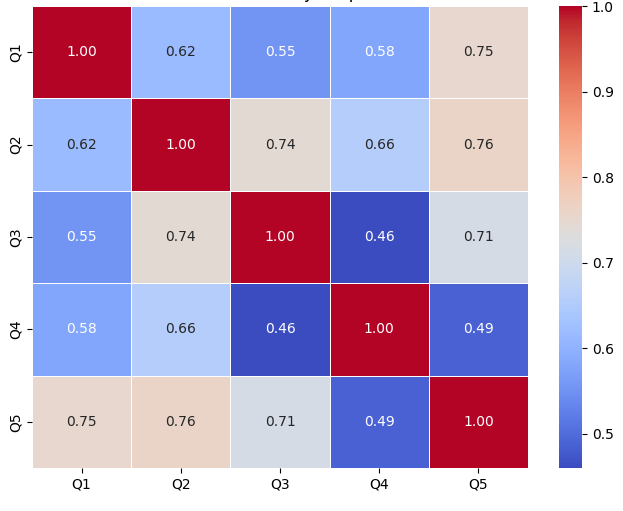}
    \caption{This figure is a heatmap showcasing the A Posteriori survey responses.}
    \label{fig:aposteriori}
\end{figure*}

The A Posteriori survey focused on participants’ self-reported familiarity with knowledge graphs and related tools. Participants responded to five items using a five-point Likert scale (1 = Strongly Disagree to 5 = Strongly Agree). The items were:

\begin{itemize}
    \item \textbf{Q1:} I understand what a knowledge graph is and how it is used.
    \item \textbf{Q2:} I have prior experience constructing or working with knowledge graphs.
    \item \textbf{Q3:} I am familiar with SPARQL, the query language used for querying knowledge graphs.
    \item \textbf{Q4:} I have used the InK Browser before to interact with knowledge graphs.
    \item \textbf{Q5:} I feel confident in applying knowledge graph principles to real-world problems.
\end{itemize}

Comprehensive knowledge of the participants' conceptual, technical, and practical expertise with semantic web technologies was intended to be captured by all of these.  

Q2 and Q5 showed the largest correlation ($r = 0.76$), indicating the confidence in using knowledge graphs outside of academic or scientific contexts is quite increased by actual use. 

A similarly high correlation was found between Q2 and Q3 ($r = 0.74$), indicating that those who had prior experience working with knowledge graphs were also more likely to be familiar with SPARQL. Since SPARQL is a standard query language for retrieving RDF data, this relationship, which is based on the co-evolution of tools and conceptual understanding, is both practical and theoretical.

In contrast, Q3 (SPARQL familiarity) and Q4 (prior use of the InK Browser) were only slightly correlated ($r = 0.46$). This relatively weaker relationship implies that SPARQL knowledge and tool-specific familiarity may emerge independently. For example, a user may learn SPARQL through coursework without ever encountering the InK Browser, or vice versa.

According to the distribution of observed correlations (0.46–0.76), the survey items measure characteristics that are linked but not redundant. This demonstrates strong credibility since the questionnaire focuses on several but related aspects of KG familiarity, including theory, practice, and tool-specific experience.

\subsection{Usability Survey Results}

\begin{figure*}[t]
    \centering
    \includegraphics[width=0.7\linewidth]{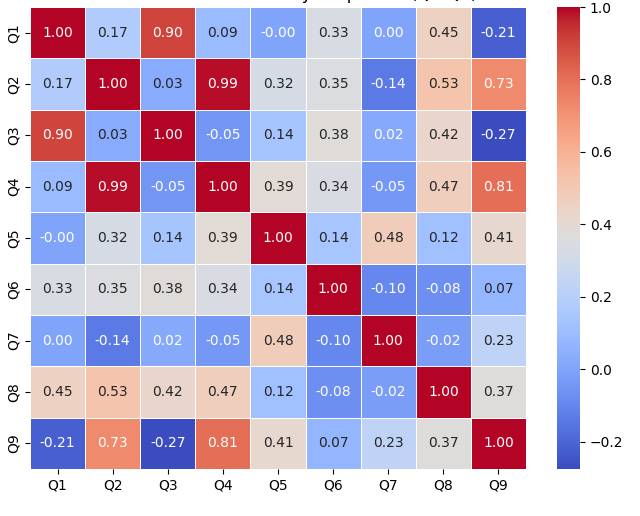}
    \caption{This figure is a heatmap showcasing the SUS survey responses.}
    \label{fig:sus}
\end{figure*}

The System Usability Scale (SUS) aimed to evaluate how participants perceived the usability of both the InK Browser and the raw RDF dataset. Participants responded to ten questions (again using a five-point Likert scale), covering ease of task completion, learning curve, need for support, and potential for future. They were also being prompted to give us a free-form comment as feedback.

The survey items were:

\begin{itemize}
    \item \textbf{Q1:} It was easy to complete query tasks using the InK Browser.
    \item \textbf{Q2:} It was easy to complete query tasks using the raw dataset.
    \item \textbf{Q3:} It was easy to find answers to task-related questions using the InK Browser.
    \item \textbf{Q4:} It was easy to find answers to task-related questions using the raw dataset.
    \item \textbf{Q5:} I would need the support of a technical person to use the tool effectively.
    \item \textbf{Q6:} I would use the InK Browser again for interacting with structured data.
    \item \textbf{Q7:} A tutorial is necessary before using the tool.
    \item \textbf{Q8:} I became familiar with how to use the InK Browser quickly.
    \item \textbf{Q9:} I became familiar with the raw dataset quickly.
    \item \textbf{Q10:} Please provide any additional comments or feedback. (Open-ended; excluded from correlation analysis)
\end{itemize}

Participants' ease of completing tasks and locating answers using the raw dataset were highly correlated, as evidenced by the highest correlation ($r = 0.99$) between Q2 and Q4. When using the InK Browser, there was a clear correlation between task execution and knowledge retrieval, as seen by the similarly high correlation ($r = 0.90$) between Q1 and Q3. 

On the other hand, satisfaction with utilizing the InK Browser was found to be negatively correlated with familiarity with the raw dataset. In this case, a negative correlation of $r = -0.27$ was found between Q3 (using the InK Browser to discover answers) and Q9 (rapid familiarity with the raw dataset). This suggests that proficiency with raw RDF is not equivalent to improved UI performance. These findings imply that, regardless of past technical knowledge of the data, the InK Browser facilitates an alternative, arguably accessible method of interaction.

Interestingly, Q1 (ease of completing tasks using the InK Browser) was uncorrelated with Q5 (need for technical support), with $r \approx 0.00$. This suggests that some participants may have found the tool easy to use but still felt insecure in their technical ability, or vice versa.

\section{Discussion}
\label{sec:discussion}

\subsection{Statistical Analysis Interpretation}

To assess the tool's accuracy and speed, we utilized both independent and paired Student's t-tests. The independent t-test tested performance differences between separate user groups, such as tool users and non-users, while the paired t-test assessed differences within the same users both before and after they used the tool. These tests can be used to identify mean differences and presume that the data is normally distributed. Furthermore, we used the Wilcoxon signed-rank test to address cases where the assumption of normalcy might not apply, including small or skewed samples. This approach supported the results under more flexible statistical assumptions, which increased the reliability of our findings.

With a mean accuracy of 3.6364 and a variation of 0.4132 for tool usage, in contrast to a mean of 1.7273 and a variance of 0.8347 for non tool usage, the data shown in Table \ref{tab:tool-usage-summary-accuracy} show a significant improvement in task accuracy. As seen in Table  \ref{tab:accuracy-test} with a T-statistic of 7.8314 and a p-value below 0.0001, an independent t-test confirms that this task accuracy difference is statistically significant, suggesting a high probability that the observed difference is not the result of chance. The paired t-test, which has a p-value of less than 0.0001 and a T-statistic of 9.2176, also confirms this assumption. The Wilcoxon test also verified the significant difference in task accuracy, producing a p-value of less than 0.0001 and a statistic of 0.0000, supporting the hypothesis that the tool positively impacted participants' performance.

In terms of task completion times, the Wilcoxon test, paired and independent t-tests as observed in \ref{tab:time-test} , both revealed that tool usage resulted in finished tasks far more quickly than raw data navigation. Both the independent t-test (T-statistic = 2.1474, p < 0.05) and the paired t-test (T-statistic = 2.1473, p < 0.05) showed that the tool's use reduced completion time, and both tests showed that the decrease was statistically significant. Additionally, this result was validated by the Wilcoxon test (statistic = 51.0000, p < 0.05). 

\subsection{A Posteriori and Usability Syrvey result interpretation}

There appears to be a strong correlation between ease of use and the capacity to find answers efficiently, as users who responded similarly to questions in the Usability questionnaire about how easy it was to use the InK Browser to complete query tasks (Q1) also tended to respond similarly to questions about how easy it was to navigate and find answers within the tool (Q3) as seen in \ref{fig:sus}. The user experience of the raw dataset and the tool was consistent, as evidenced by the fact that those who considered the raw dataset easy to use (Q2) also reported finding it easier to navigate (Q4) and discover answers (Q9). According to these relationships, enhancing a specific component of the tool's usability\-like task navigation may result in improvements in other areas as well.

Further review of the comments identifies potential areas for improving the user experience. One participant emphasized that users should be introduced to the raw data format and given a quick rundown of the tool's features, especially its pop-up windows. This could avoid confusion and make users familiar with the interface. Respondents who found the InK Browser easy to use (Q1, Q2) were more likely to find the tool useful for completing tasks and interacting with the dataset, which is consistent with the correlation analysis. The results highlight how the overall user experience might be greatly improved by making the tool easier to use and by offering improved data engagement.

Strong connections exist between the confidence users have in using knowledge graph principles, their expertise with knowledge graphs, and their experience with SPARQL, according to the correlation matrix and a posteriori answers investigation as seen in \ref{fig:aposteriori}.   The likelihood that respondents would report having experience creating or working with knowledge graphs (Q2) and feeling comfortable implementing knowledge graph principles in real-life circumstances (Q5) was higher among those who indicated a high understanding of knowledge graphs (Q1). People who have a better idea of knowledge graphs are also more confident in their ability to use them in real-world situations, as seen by the high correlations found between Q1 and Q2 and Q1 and Q5.

\section{Conclusion}
\label{sec:conc}

The purpose of this study was to assess the usefulness of the InK Browser, a tool made to help users navigate and comprehend knowledge graphs (KGs). The study specifically focused on exploring how increasing the size of a knowledge graph and adding additional participants could further demonstrate the tool's capacity to help in the understanding of large and complex datasets. The study also gave the chance to evaluate the InK Browser's power in a realistic and data-rich setting by utilizing the CoreScholar dataset, which shows highly complicated relationships between researchers, publications, and participants.

Subjects were prompted to answer eight questions separated into two parts, four while going through raw data and four while navigating the Browser. The order of that was random and was alternating between subjects to avoid any bias. The only data that was collected was the participants' answers to questions regarding the dataset and time to complete TaskA and TaskB,each  containing  four questions respectively. Once that data was organized for analysis, statistical tests like independent and paired student t-test  and Wilcoxon  tests we performed confirming our original hypotheses, that the use of the InK Browser will result in both faster (Hypothesis 1) and more accurate (Hypothesis 2) task completion.

Analysis of the SUS and a posteriori replies demonstrates how effective the InK Browser tool is, especially when it comes to task accuracy, completion time, and satisfaction among users. Participants who had previously used SPARQL and understood knowledge graphs performed better, using the tool to complete tasks with greater accuracy and rapidly.   The tool's important completion time reduction suggests that it also improves user efficiency. The positive relationships between prior knowledge and job completion ease indicated that the tool was well-received, particularly by those with a greater familiarity with knowledge graphs.

\subsection*{Future Work}
The InK Browser is currently being evolved with the following functionalities:

\begin{enumerate}
    \item Stream multimedia like tutorial videos regarding KG educational materials.

    \item Aiming for the user to be able to watch an educational video and when pausing it to be able to acquire all the information regarding a specific aspect of the tutorial while being prompted for more information.

    \item Ability to extent the browser even more, allowing personalized learning paths to be attached for every user, walking them through a sequence of learning steps.

    \item Adress participant's feedback regarding any functionality improvements.

\end{enumerate}
\medskip

\noindent\emph{Acknowledgment.} The author(s) acknowledge support from the National Science Foundation under award \#2333532 ``Proto-OKN Theme 3: An Education Gateway for the Proto-OKN (EduGate). Any opinions, findings, and conclusions or recommendations expressed in this material are those of the author(s) and do not necessarily reflect the views of National Science Foundation or the U.S. Department of Education.


\bibliographystyle{abbrv}
\bibliography{refs}

\end{document}